\documentclass[fleqn,twoside,twocolumn,nofootinbib]{revtex4} 
\usepackage{ujp} 
\begin{document}
\title[SPECIFIC FEATURES OF MOTION OF CATIONS AND ANIONS]
{SPECIFIC FEATURES OF MOTION OF CATIONS\\ AND ANIONS IN ELECTROLYTE SOLUTIONS}%
\author{L.A. Bulavin}
\affiliation{Taras Shevchenko National University of Kyiv}
\address{2/1, Prosp. Academician Glushkov, Kyiv 03127, Ukraine}
\email{bulavin@univ.kiev.ua}
\author{I.V. Zhyganiuk}%
\affiliation{I.I. Mechnikov Odesa National University}
\address{2, Dvoryans'ka Str., Odesa 65026, Ukraine }
\author{M.P. Malomuzh}%
\affiliation{I.I. Mechnikov Odesa National University}
\address{2, Dvoryans'ka Str., Odesa 65026, Ukraine }
\email{mnp@normaplus.com}
\author{K.M.~Pankratov}%
\affiliation{Odesa National Polytechnic University}%
\address{1, Shevchenko Str., Odesa 65044, Ukraine}%

\udk{532} \pacs{66.10.C} \razd{\secvi}

\setcounter{page}{893}%
\maketitle

\begin{abstract}
The nature of mobility of ions and water molecules  in dilute aqueous
solutions of electrolytes (at most fifteen water molecules per ion)
is investigated. It is shown that the behavior of the mobility
coefficients of water molecules and ions, as well as the
self-diffusion coefficients of water molecules, are determined by the
radii of their hard shells rather than by the effect of the hydrogen
bond network. It is established that the influence of hydration
effects on the density of the system and the self-diffusion
coefficients of water molecules does not exceed several per cent.
Based on microscopic concepts, it is shown that the different
behaviors of a $\rm K^{+}$ cation and an $\rm F^{-}$ anion with equal
rigid radii are in good agreement with specific features of
the intermolecular interaction described by the generalized
Stillinger--David potential \cite{1,2}.
\end{abstract}%


\begin{table}[b]
\noindent\caption{Self-diffusion coefficients of water molecules in
water solutions of singly charged electrolytes \cite{5}
}\vskip3mm\tabcolsep2pt

\noindent{\footnotesize\begin{tabular}{c c c c c c c c c}
 \hline \multicolumn{1}{c}
{\rule{0pt}{9pt}$ $} & \multicolumn{1}{|c}{$ $}&
\multicolumn{1}{|c}{$ $}& \multicolumn{1}{|c}{$\rm LiBr(30)$}&
\multicolumn{1}{|c}{$\rm LiI(24.8)$}\\%
\hline%
$D_s^{(w)}\times 10^5,~ \mathrm{cm^2/s}$& & & 2.3 & 2.18 \\%
\hline
{\rule{0pt}{9pt}$ $}& \multicolumn{1}{|c} {}& \multicolumn{1}{|c}{$\rm NaCl(15.9)$}& \multicolumn{1}{|c} {$\rm NaBr (16.5)$}& \multicolumn{1}{|c} {$\rm NaI (16.1)$}\\%
\hline
$D_s^{(w)}\times 10^5, ~\mathrm{cm^2/s}$&  & 2.14 & 2.26 & 2.38 \\%
\hline
{\rule{0pt}{9pt}$ $}& \multicolumn{1}{|c} {$\rm KF(15)$}& \multicolumn{1}{|c}{$\rm KCl(16.1)$}& \multicolumn{1}{|c} {$\rm KBr (16.1)$}& \multicolumn{1}{|c} {$\rm KI (16.4)$}\\%
\hline
$D_s^{(w)}\times 10^5, ~\mathrm{cm^2/s}$& 1.99 & 2.44 & 2.68 & 2.8 \\%
\hline

\end{tabular}}
\end{table}

\section{Introduction}

For today, a substantial body of information is accumulated  on the
thermal motion of water molecules and ions in water electrolyte
solutions (see \cite{3,4,5,6,7,8,9,10,11,12,13}). Now, of interest
is the experimental study of the kinetic and electrophysical properties
of electrolytes. The specific features of motion of water molecules
and ions have a certain effect on the thermodynamic properties of
liquids and liquid systems \cite{14}. The study of a structure and
a character of thermal molecular motion in electrolytes with the help
of the quasielastic scattering of slow neutrons \cite{5,6,7,8,9,10} and
molecular dynamics methods \cite{11,12,13} acquires still larger
importance.

There also exists a large number of substantial theoretical  works
\cite{15,16} devoted to the description of properties of
electrolytes. The main efforts in them are concentrated on the
adequate allowance for hydration effects and the correct reflection of
the role of hydrogen bonds. However, a lot of questions (very simple,
at first sight) remain practically unanswered. A typical example is
peculiarities of the manifestation of hydration effects depending on the
radii of ions and the signs of their charges.

Let us illustrate this fact by the example concerning the behavior
of the  self-diffusion coefficients of water molecules as functions
of the cation and anion sizes. The corresponding values of the
self-diffusion coefficients $D_s^{(w)}$ of water molecules in
several dilute solutions of electrolytes at the temperature $T=296$
K are given in Table 1 presenting the dependence of $D_s^{(w)}$ on
the anion size. The solution concentration in the table is specified
by the number $z_{w}$ of water molecules per one ion (the value of
$z_{w}$ is given in parentheses near the chemical formula of an
electrolyte, for example $\rm NaCl (15{.}9)$). The minimal value
$z_{w}=15$ thus corresponds to the electrolyte concentration equal
to $3{.}3$ mole 
mutual effect of cations and anions can be neglected to a
satisfactory accuracy. It is also worth adding to the table the
self-diffusion coefficient of water molecules in the $\rm CsI
(17{.}4)$ solution: $D_{s}^{(w)}$=2{.}71$\times 10^5$ cm$^2$/s.

We note that, according to \cite{17,18}, the
self-diffusion  coefficient of molecules in water at the same
temperature $T=296$ K is equal to
\begin{equation} \label{Ds}
D_s^{(w)}=2{.}35\times 10^{-5} {\mathrm{cm}^2/\mathrm{s}}.
\end{equation}

The number of hydrogen bonds formed by one molecule in water at the
same temperature is approximately equal to $n_{\rm H}\approx 2{.}5$
\cite{19,20,21,22}. In this case, the average distance between water
molecules is close to the hydrogen bond length $l_{\rm H}\approx
2{.}8 $ \AA. Comparing (\ref{Ds}) with the values of the
self-diffusion coefficients given in Table 1, one can conclude that
the self-diffusion coefficients of water molecules considerably
depend on the combinations of cations and anions even in dilute
solutions of electrolytes.

In this work, we analyze the mobility coefficients  of water
molecules as functions of the cation and anion sizes in the close
connection with the behavior of the solution density. In this way,
we try to make certain conclusions concerning the role of
hydration effects in solutions of electrolytes. In addition, to
discover the nature of asymmetry effects in the interaction of
anions and cations with water molecules, the microscopic ideas about the
character of their interaction are involved. This aspect of the
problem is solved on the basis of the generalized potential of
interparticle interaction proposed in \cite{1,2,23}.

\section{Dependence of the Self-Diffusion and Mobility Coefficients of
 Water Molecules and the Mobility Coefficients of Ions on Their Size}

To analyze the results presented in Table 1, let  us consider how
the self-diffusion coefficients of water molecules in solutions of
electrolytes correlate with the radii of dissolved ions. The latter
are not determined unambiguously, but considerably depend on the
method of their determination. In this connection, we consider the
following ionic radii: 1) those found from crystallographic
conceptions; 2) those chosen to favor the correct reproduction
of the molecular dynamics in solutions of electrolytes by means of
computer simulation, and 3) those estimated from the ion
polarization. To make the picture complete, we consider also
the ionic radii determined from hydrodynamic considerations. The
corresponding results are gathered in Table 2.


\begin{table}[b]
\noindent\caption{Rigid (three upper rows) and Stokes  radii of
cations and anions}\vskip3mm\tabcolsep5pt

\noindent{\footnotesize\begin{tabular}{c c c c c c c c c}
 \hline \multicolumn{1}{c}
{\rule{0pt}{9pt}}& \multicolumn{1}{|c}{Li$^{+}$} &
\multicolumn{1}{|c}{Na$^{+}$} & \multicolumn{1}{|c}{K$^{+}$}&
\multicolumn{1}{|c}{Cs$^{+}$}& \multicolumn{1}{|c}{F$^{-}$}&
\multicolumn{1}{|c}{Cl$^{-}$}& \multicolumn{1}{|c}{Br$^{-}$} &
\multicolumn{1}{|c}{I$^{-}$}\\%
\hline%
 $r_{\rm c}$, {\AA}&0.6& 0.95& 1.33&1.69  & 1.36& 1.81 & 1.95 & 2.16 \\%
$r_{\sigma}$, {\AA} &0.76& 1.3& 1.67& 1.94& 1.56& 2.2 & 2.27 & 2.59 \\%
$r_{\alpha}$, {\AA} &0.45& 1.12& 1.41& 2.02& 1.51& 2.33 & 2.55 & 2.93 \\%
$r_s^{(\mu)}$, {\AA}&2.38& 1.84& 1.25&1.19 & 1.66& 1.21 & 1.18 & 1.19 \\%
$r_{\rm c}^{(D)}$, {\AA}&1.91& 1.88& 1.14&1.15 & 1.77& 1.33 & 1.22 & 1.35 \\%
\hline
\end{tabular}}
\end{table}

In the first row of Table 2, one can see the crystallographic radii
$r_{\rm c}$ \cite{24}. The second row contains the ionic radii
$r_{\sigma}$ found from computer experiments \cite{25} for
the description of the dispersion (van der Waals) interaction between
ions and water molecules. The ionic radii $r_{\alpha}$ determined
from the ion polarizations $\alpha$ (see Table 3) according to the
formula
\begin{equation}\label{pol}
r_{\alpha}=1.5\alpha^{1/3}
\end{equation}
are presented in the third row. It is worth noting  that, in the
simplest model of an ion presenting it as a conducting sphere, the
polarization is equal to $\alpha=r_{\mathrm{sph}}^3$, where
$r_{\mathrm{sph}}$ is the sphere radius. With regard for the
fact that a conducting sphere reflects ion properties only
approximately, the coefficient $1{.}5$ in (\ref{pol}) is chosen to
provide the minimum deviation of the ionic radii $r_{\alpha}$ from
$r_{\rm c}$. The radii $r_{\rm c}$, $r_{\sigma}$, and $r_{\alpha}$
will be called rigid ionic radii.


\begin{table}[b]
\noindent\caption{Ion polarizations, \AA\boldmath$^3$ \cite{26}
}\vskip3mm\tabcolsep7.2pt

\noindent{\footnotesize\begin{tabular}{c c c c c c c c c}
 \hline \multicolumn{1}{c}
{\rule{0pt}{9pt}Li$^{+}$} & \multicolumn{1}{|c}{$\mathrm{Na}^{+}$} &
\multicolumn{1}{|c}{$\mathrm{K}^{+}$}&
\multicolumn{1}{|c}{$\mathrm{Cs}^{+}$ }&
\multicolumn{1}{|c}{F$^{-}$}& \multicolumn{1}{|c}{$\mathrm{Cl}^{-}$}
& \multicolumn{1}{|c}{$\mathrm{Br}^{-} $} &
\multicolumn{1}{|c}{$\mathrm{I}^{-}$}\\%
\hline%
0{.}029& 0{.}179& 0{.}83& 2{.}42& 1{.}04& 3{.}66 & 4{.}77 & 7{.}1 \\%
\hline
\end{tabular}}
\end{table}

For the sake of comparison, the fourth and fifth rows  of Table 2
present the Stokes ionic radii determined from the formulas
\begin{equation}
r_s^{(\mu)}=\frac{1}{6\pi\eta{\mu}^{(I)}}
\end{equation}
and
\[
r_s^{(D)}=\frac{k_{\rm B} T}{6\pi\eta D_s^{(I)}},
\]
where ${\mu}^{(I)}$ and $ D_s^{(I)}$ stand for the mobility and
the self-diffusion coefficient of ions determined by computer modeling
in \cite{25,26,27}, whereas $\eta$ is the shear dynamic viscosity of
electrolyte solutions \cite{28}.

An important supplement to these data can be the Stokes radii of water
molecules in electrolyte solutions gathered in Table 1. They are
determined by the formula
\begin{equation}
r_s^{(w)}=\frac{k_{\rm B} T}{6\pi\eta D_s^{(w)}}
\end{equation}
and form Table 4 as well. In contrast to $r_s^{(\mu)}$  and
$r_s^{(D)}$, the values of $r_s^{(w)}$ practically do not depend on
the type of a dissolved electrolyte, as is worth expecting. At the
same time, the obtained radii $r_s^{(w)}$ are noticeably smaller
than the hard-sphere radius of water molecules, if identifying the
latter with a half of the hydrogen bond length \cite{29,30}. Here,
it is worth noting that hydrodynamic concepts can be used only on the
scales significantly exceeding the corresponding molecular ones.
That is why we make conclusion that the application of
hydrodynamic conceptions to the self-diffusion process of water
molecules is inconsistent.


\begin{table}[b]
\noindent\caption{Stokes radii of water molecules in solutions  of
electrolytes gathered in Table 1 }\vskip2mm\tabcolsep4pt

\noindent{\footnotesize\begin{tabular}{c c c c c c c c c}
 \hline \multicolumn{1}{c}
{\rule{0pt}{9pt}} & \multicolumn{1}{|c}{LiBr}&
\multicolumn{1}{|c}{LiI}& \multicolumn{1}{|c}{NaCl}&
\multicolumn{1}{|c}{NaI}& \multicolumn{1}{|c}{KF}&
\multicolumn{1}{|c}{KCl} & \multicolumn{1}{|c}{KBr} & \multicolumn{1}{|c}{KI}\\%
\hline%
$\eta $& 1{.}009& 0{.}97& 0{.}997& 0{.}957& 1{.}053 & 0{.}936& 0{.}917& 0{.}912 \\%
$r_s^{(w)}$ &0{.}934&1{.}025& 1{.}015 & 0{.}951 & 1{.}034 & 0{.}949 & 0{.}881 & 0{.}85\\%
\hline
\end{tabular}}
\end{table}

One can see that the ionic radii used in the computer  experiment
somewhat exceed their crystallographic values. In the both cases,
one observes, however, a monotonous growth of the cation and anion
radii with increase in masses. The same dependence is registered for
the ionic radii obtained from the ion polarizations.

On the contrary, the Stokes radii of water ions and molecules
demonstrate the opposite behavior. The largest Stokes radius
corresponds to a small $\rm Li^{+}$ cation, whereas the least one
is observed for an $\rm I^{-}$ anion. There also exists a certain
asymmetry in the behavior of cations and anions of the same radius.
Indeed, the differences in the Stokes radii of $\rm K^{+}$ and $\rm
F^{-}$ ions, as well as in their mobilities (see Table 5),
significantly exceed the differences in the values of their rigid
radii $r_{\rm c}$, $r_{\sigma}$, and $r_{\alpha}$. This fact is to a
certain extent unexpected. It should be related to the noticeably
different character of the interaction of water molecules with $\rm
K^{+}$ and $\rm F^{-}$ ions, as well as with all other cations and
anions. This important fact is thoroughly analyzed in Section 4 by
the example of $\rm K^{+}$ and $\rm F^{-}$.


\begin{table}[b]
\noindent\caption{Mobility coefficients of ions determined  from the
conductivity of electrolytes (\boldmath$\mu_{\rm
c}={\tilde{\mu}}_{\rm c}\times 10^8$) \cite{27} and by computer
simulation of the mean-square shift of an ion
(\boldmath$\mu_{\Gamma}={\tilde{\mu}}_{\Gamma}\times 10^8$) and its
autocorrelation velocity function
(\boldmath$\mu_{\upsilon}={\tilde{\mu}}_{\upsilon}\times 10^8$)
\cite{25}}\vskip3mm\tabcolsep6.1pt

\noindent{\footnotesize\begin{tabular}{c c c c c c c c c}
 \hline \multicolumn{1}{c}
{\rule{0pt}{9pt}}& \multicolumn{1}{|c}{Li$^{+}$} &
\multicolumn{1}{|c}{$\mathrm{Na}^{+}$} &
\multicolumn{1}{|c}{$\mathrm{K}^{+}$}&
\multicolumn{1}{|c}{$\mathrm{Cs}^{+}$ }&
\multicolumn{1}{|c}{$\mathrm{F}^{-}$}&
\multicolumn{1}{|c}{$\mathrm{Cl}^{-}$} &
\multicolumn{1}{|c}{$\mathrm{Br}^{-}$} &
\multicolumn{1}{|c}{$\mathrm{I}^{-}$}\\%
\hline%
 $\tilde{\mu}_{\rm c}$ &4{.}01& 5{.}19& 7{.}62&  & 5{.}7& 7{.}91 & 8{.}13 & 7{.}96 \\%
$\tilde{\mu}_{\Gamma}$ &4{.}75& 4{.}98& 7{.}12& 7{.}32& 4{.}04& 6{.}88 & 7{.}2 & 6{.}23 \\%
$\tilde{\mu}_{\upsilon}$ &4{.}59& 5{.}02& 7{.}20& 7{.}36& 3{.}85& 6{.}42 & 6{.}85 & 6{.}27 \\%
\hline
\end{tabular}}
\end{table}

As follows from Tables 1 and 2, the self-diffusion  coefficients of
water molecules as functions of their rigid radii satisfy the
following regularities:

1) for dilute lithium and sodium solutions of electrolytes,  in
which $r_{\rm c}<\frac{ 1}{ 2}\,l_{\rm H}$, the following inequality
is met: $D_s^{(w)}({\rm el})<D_s^{(w)}$. The condition $r_{\rm
c}<\frac{ 1}{ 2}\,l_{\rm H}$ is evidently violated only for $\rm
Cs^{+}$. In solutions of potassium electrolytes, where $r_{\rm c}
({\rm K^+})\sim\frac{ 1}{ 2}\,l_{\rm H}$, one observes a transition
from the previous relation between the self-diffusion coefficients
of water molecules to the inequality
$D_s^{(w)}(\mathrm{el})>D_s^{(w)}$;

2) in solutions of electrolytes with a fixed cation, except  for
lithium ones, the self-diffusion coefficients of water molecules
grow with the anion radius;

3) in lithium electrolytes, the character of the  dependence of
$D_s^{(w)}(\mathrm{el})$ on the anion radius is opposite to the
second conclusion.

With regard for the fact that the concentrations of different singly
charged cations and anions are close to each other, the differences
in the behavior of $D_s^{(w)}(\mathrm{el})$ can be caused by the
following reasons: 1) geometrical factors and 2) their different
effect on the structure of the local environment (hydration
effects). The first possibility should be discarded, as the geometrical
obstacles must decrease with the cation radius, which conflicts
with experimental data. But a reconfiguration of the local
structure of water in the direct environment of cations and anions
appears more noticeable, as their rigid radii increase.

Indeed, comparing the values of $r_s^{(\mu)}$, $r_s^{(D)}$, and
$r_s^{(w)}$, one can  see that, as the rigid radii of
cations and anions increase, their Stokes radii approach the Stokes radius of
water molecules. It is natural to interpret this fact as the joint
drift of large ions together with water molecules that enter
hydrated shells formed around them. It is worth noting that the
effect of a local reconfiguration of the solution structure is
insignificant, because the rise or fall of the self-diffusion
coefficient of water molecules usually does not exceed ten per cent
and is proportional to the mole concentration of electrolyte
impurities.

\section{Analysis of the Hydration Energy of Ions and the Density of Electrolyte Solutions}

The dependence of the Stokes ionic radii  on the rigid sizes of
cations and anions is closely related to the phenomenon of
the so-called positive or negative ion hydration \cite{14,31,32}.
Indeed, according to \cite{33}, the sign of the ion hydration energy
is determined by the quantity $\Delta W=W-W_{\rm I}$, where $W$ is
the average binding energy of two neighboring water molecules, and
$W_{\rm I}$ is the average value of their interaction energy in the
presence of ions. An ion close to the neighboring water
molecules results in their additional polarization and an increase
of the attractive forces acting between them. In the absence of
hydrogen bonds, the difference $W-W_{\rm I}$ must be positive, as
the both terms $W$ and $W_{\rm I}$ are negative. If a cation or
an anion partially destroys hydrogen bonds, then the absolute value
of $W_{\rm I}$ can become lower than that of $W$, i.e. the negative
sign of $\Delta W$ will testify to the local breakage of the
hydrogen bond network. Table 6 presents the dimensionless
combination $\Delta{\tilde W}=\Delta W/3k_{\rm B} T_m$ (where
$3k_{\rm B} T_m$ is the energy of translational thermal motion of
two water molecules at the crystallization temperature $T_m=273$ K)
as a function of the type of ions. Analyzing Table 6, one can see
that $\rm Li^{+}$ and $\rm Na^{+}$ cations, whose rigid radii are
less than a half of the hydrogen bond length (i.e. a half of the
average intermolecular distance between water molecules), conserve a
rather intact structure of the local hydrogen bond network and
intensify the interaction between neighboring molecules. A
relatively small value of $\Delta W(\rm Na^{+})$ can be interpreted
as a consequence of the more significant bending of hydrogen bonds
as compared to that as a reaction to the introduction of a small-sized
$\rm Li^{+}$ cation. A $\rm K^{+}$ cation, whose rigid radius is
close to a half of the hydrogen bond length, causes a damage of the
local hydrogen bond network, which results in the negative sign of
$\Delta W$. In the same way, one can explain the negative sign of
$\Delta W$ for a $\rm Cl^{-}$ anion.


\begin{table}[b]
\noindent\caption{Dependence of \boldmath$\Delta{\tilde W}$ on the
type of ions
 \cite{33}} \vskip3mm\tabcolsep12.7pt

\noindent{\footnotesize\begin{tabular}{c c c c c c c c c}
 \hline \multicolumn{1}{c}
{\rule{0pt}{9pt}$$} & \multicolumn{1}{|c}{$\mathrm{Li}^{+}$}&
\multicolumn{1}{|c}{$\mathrm{Na}^{+}$}& \multicolumn{1}{|c}{$
\mathrm{K}^{+}$}& \multicolumn{1}{|c}{$\mathrm{F}^{-}$}&
\multicolumn{1}{|c}{$\mathrm{Cl}^{-}$}\\%
\hline%
$\Delta\tilde{W}$& 3{.}4& 0{.}06& -4{.}6& 0{.}52& -2{.}0 \\%
\hline
\end{tabular}}
\end{table}

At the same time, it is worth noting a significant  asymmetry in the
behavior of $\Delta W$ for $\rm K^{+}$ and $\rm F^{-}$ ions having
almost identical rigid radii (see Table 2). The positive sign of
$\Delta W(\rm F^{-})$ testifies to the fact that an $\rm F^{-}$ anion
does not completely break the local configuration of hydrogen bonds,
though significantly deforms it. It is clearly proved by a
relatively small value of $\Delta{\tilde W}$. To our opinion, the
appearance of this asymmetry is caused by the polarization component
of the interaction between ions and water molecules. The existence of
the asymmetry in the behavior of cations and anions was first noted
in \cite{34,35}.

On the other hand, the role of hydration effects can be estimated, by
analyzing the density of weak solutions of electrolytes (see
Table~7). We must actually estimate a deviation of the density of
the real solution from that of the ideal one. By definition, an
electrolyte solution can be considered as ideal if the volumes
occupied in it by water molecules and dissolved ions acquire the
same values as in water and electrolyte melts. For such an
electrolyte solution, the following relation must be met:
\begin{equation} \label{ups_el}
\upsilon_{\rm w}^{(0)} n_{\rm w}+ (\upsilon_{\rm
c}^{(0)}+\upsilon_{\rm a}^{(0)})n_{\mathrm{el}}=1.
\end{equation}
Here, $n_\mathrm{w}$ and $n_{el}$ denote the number densities of
water and electrolyte molecules in the real solution, while
$\upsilon_{\rm w}^{(0)}$, $\upsilon_{\rm c}^{(0)}$, and
$\upsilon_{\rm a}^{(0)}$ are the volumes occupied by water molecules
and ions in water and in melted or solid electrolytes. The values of
$n_\mathrm{w}$ and $n_{\mathrm{el}}$ can be obtained from the
definition of the mass density of an electrolyte solution:
\[
\rho=m_\mathrm{w} n_\mathrm{w}+(m_{\rm c}+m_{\rm a}) n_{\mathrm{el}}
\]
and its weight concentration
\[
x=\frac{(m_{\rm c}+m_{\rm a}) n_{\mathrm{el}}}{m_\mathrm{w}
n_\mathrm{w} +(m_{\rm c}+m_{\rm a})n_{\mathrm{el}}}.
\]
Hence,
\begin{equation} \label{nwel}
n_\mathrm{w}=(1-x)\frac{\rho}{m_\mathrm{w}}, \quad n_{\mathrm{el}}=x
\frac{\rho}{m_\mathrm{c}+m_\mathrm{a}}.
\end{equation}
A real electrolyte solution is described by the relation similar to (\ref{ups_el}):
\begin{equation} \label{ratushnyak}
\upsilon_{\rm w} n_{\rm w}+(\upsilon_{\rm c}+\upsilon_{\rm
a})n_{\mathrm{el}}=1,
\end{equation}
though it contains the real volumes occupied by a water molecule,
cation,  and anion. That is why a deviation of the dimensionless
combination
\begin{equation} \label{knoprst}
\delta=\upsilon_{\rm w}^{(0)} n_{\rm w}+ (\upsilon_{\rm
c}^{(0)}+\upsilon_{\rm a}^{(0)})n_{\mathrm{el}}-1
\end{equation}
from zero serves as the measure of adeviation of a water electrolyte
solution from the ideal one. Moreover, positive values of the
non-ideality parameter ($\delta>0$) will correspond to the formation
of hydrated shells around ions, in which the water density will be
larger as compared to the density of bulk water. Indeed, the
formation of a hydrated shell is accompanied by a decrease of the
volume occupied by a water molecule
($\upsilon_\mathrm{w}<\upsilon_{\rm w}^{(0)}$). The variation of the
volumes occupied by cations and anions in weak electrolyte solutions
can be neglected.


\begin{table}[b]
\noindent\caption{Densities of water electrolyte solutions at the
fixed concentration \boldmath$x_{\mathrm{el}}=4$ wt.}
\vskip3mm\tabcolsep15.2pt

\noindent{\footnotesize\begin{tabular}{c c c c c c c c c}
 \hline \multicolumn{1}{c}
{\rule{0pt}{9pt}} & \multicolumn{1}{|c}{$\mathrm{F}^{-}$}&
\multicolumn{1}{|c}{$\mathrm{Cl}^{-}$}& \multicolumn{1}{|c}{$
\mathrm{Br}^{-}$}&
\multicolumn{1}{|c}{$\mathrm{I}^{-}$}\\%
\hline%
$\mathrm{Li}^{+}$& & --0{.}001&  &  \\%
$\mathrm{Na}^{+}$& & 0{.}003 &  & \\%
$\mathrm{K}^{+}$&  & & 0{.}003 & 0{.}004 \\%
$\mathrm{Cs}^{+}$&  & 0{.}005 & & \\%
\hline
\end{tabular}}
\end{table}

Let us substitute the following values of the
parameters $\upsilon_{\rm w}^{(0)}$, $\upsilon_\mathrm{c}^{(0)}$,
and $\upsilon_{\rm a}^{(0)}$ in (\ref{knoprst}): $\upsilon_{\rm
w}^{(0)}=\frac{\displaystyle m_\mathrm{w}}{\displaystyle \rho_{\rm
w}^{(0)}}$,
$\upsilon_\mathrm{c}^{(0)}+\upsilon_\mathrm{a}^{(0)}=\frac{\displaystyle
m_\mathrm{c}+m_\mathrm{a}}{\displaystyle \rho_{el}^{(0)}}$, where
$\rho_{\rm w}^{(0)}$ and $\rho_{\mathrm{el}}^{(0)}$ are the water
and electrolyte densities in the liquid and amorphous states. With
regard for the number densities of water and electrolyte
impurities determined by relations (\ref{nwel}), Eq. (\ref{knoprst})
takes the form
\begin{equation}\label{delta1}
\delta=\frac{\rho}{\rho_{\rm w}^{(0)}}\left[1+x\left(\frac{\rho_{\rm
w}^{(0)}}{\rho_{\mathrm{el}}^{(0)}}-1 \right)\right].
\end{equation}
The non-ideality parameter of a dilute solution can be interpreted
in somewhat another way. Taking into account that the dominant
contribution to $\delta$ is given by a variation of the volume
occupied by one water molecule, Eqs. (\ref{knoprst}) and
(\ref{ratushnyak}) yield: $\delta\approx n_{\rm w} \delta
\upsilon_{\rm w}$. So, to an acceptable accuracy, the non-ideality
parameter for a dilute solution is
\begin{equation}
\delta\approx\frac{\delta \upsilon_{\rm w}}{\upsilon_{\rm w}^{(0)}}.
\end{equation}
The densities of some electrolyte solutions, as well as the
parameters of their non-ideality $\delta,$ are given in Tables 7 and
8. It is worth noting that the non-ideality parameters obtained
according to formula (\ref{delta1}) are averaged over the number of
water molecules surrounding a certain ion. The change of the
relative volume of water molecules in the first hydrated layer of an
ion is larger by a factor of $z_{\rm w}/z_1$, where $z_1$ is the
number of water molecules getting to the first hydrated layer.


\begin{table}[b]
\noindent\caption{Non-ideality parameters of the
 solutions at the fixed concentration \boldmath$x_{\mathrm{el}}=4$ wt.}
\vskip3mm\tabcolsep15.2pt

\noindent{\footnotesize\begin{tabular}{c c c c c c c c c}
 \hline \multicolumn{1}{c}
{\rule{0pt}{9pt}} & \multicolumn{1}{|c}{$\mathrm{F}^{-}$}&
\multicolumn{1}{|c}{$\mathrm{Cl}^{-}$}& \multicolumn{1}{|c}{$
\mathrm{Br}^{-}$}&
\multicolumn{1}{|c}{$\mathrm{I}^{-}$}\\%
\hline%
$\mathrm{Li}^{+}$& & --0{.}001&  &  \\%
$\mathrm{Na}^{+}$& & 0{.}003 &  & \\%
$\mathrm{K}^{+}$&  & & 0{.}003 & 0{.}004 \\%
$\mathrm{Cs}^{+}$&  & 0{.}005 & & \\%
\hline
\end{tabular}}
\end{table}

In particular, the non-ideality parameter for a $\rm NaCl$
solution at $x_{\mathrm{el}}=4$ wt.
reaches $\delta\approx 0{.}003$. One can see that the relatively small values
of mobility of lithium cations and self-diffusion coefficients of water
molecules in lithium electrolytes correspond to negative non-ideality parameters.
This testifies to the fact that lithium cations do not favor the formation of
hydrated shells around them with densities exceeding that of water. In other cases,
such hydrated shells are formed. At the same time, the smallness of $\delta$ is an evidence
of the fact that hydration effects in dilute electrolyte solutions have a weak effect on their densities.

At relatively small concentrations of electrolytes, the deviation of
a solution from ideality is proportional to its mole concentration
$c=\frac{\displaystyle m_\mathrm{w}}{\displaystyle
m_\mathrm{c}+m_\mathrm{a}} x$:
\[
\delta\approx\varepsilon\, c.
\]
The quantity $\varepsilon$ also can be interpreted as a measure of
non-ideality of a solution. For the above-considered solutions, it
lies in the range $0{.}1<\varepsilon<0{.}5$.

The result described in two previous sections allow one  to conclude
that lithium cations do not radically change the local structure of
surrounding water formed with the help of hydrogen bonds. As a
result, their thermal motion can be considered as a drift in
temporary ``cavities'', whose generation is favored by the local
hydrogen bond network. As the characteristic time of such drift, one
can consider the time of existence of hexagonal rings in water, which
should be identified, according to \cite{30}, with the duration of
the settled life of water molecules $\tau_0$. The characteristic shift
of a lithium cation can be estimated by the formula
$|\Delta\textbf{r}({\rm Li^{+}})|\sim\sqrt{6{D_s}({\rm
Li^{+}})\tau_0}$, or $|\Delta\textbf{r}({\rm
Li^{+}})|\approx\upsilon_T({\rm Li^{+}})\tau_0$. The value of
$\tau_0$ at the temperature $T=295$ K is approximately equal to
$(0{.}8 \div 1)\times 10^{-12}$\, s \cite{18}.  The self-diffusion
coefficient of a $\rm Li^{+}$ cation is estimated with the use of the formula
$D_s({\rm Li^{+}}) = k_{\rm B} T {\tilde{\mu}}_{\rm c}\times 10^8 $.
As a result, the characteristic shift $|\Delta\textbf{r}({\rm
Li^{+}})|$ of a lithium cation during an elementary diffusion act is
equal to $|\Delta\textbf{r}({\rm Li^{+}})| \approx 2{.}8\times
10^{-8}$ cm. The same order of magnitude is characteristic of the
combination $\upsilon_T({\rm Li^{+}})\tau_0$. This allows one to
conclude that $|\Delta\textbf{r}({\rm Li^{+}})|$ practically
coincides with the average distance between neighboring water
molecules or the size of a hexagonal ring. A similar conclusion
can be also valid as regards the thermal motion of a ${\rm Na^{+}}$
cation. Indeed, $\left. |\Delta\textbf{r}({\rm  Na^{+}})| \right/
|\Delta\textbf{r}({\rm Li^{+}})| \sim \sqrt{\left.
{\tilde{\mu}}_{\rm c} {({\rm  Na^{+}})}
 \right/ {\tilde{\mu}}_{\rm c} {({\rm Li^{+}})} }  \sim 1{.}1$.

On the other hand, cations and anions, whose rigid radii exceed  a
half of the hydrogen bond length ($r_{\rm I}>\frac{ 1}{ 2}\, l_{\rm
H}$), destroy the hydrogen bond network around themselves and thus
move freer. The rigid radii of ${\rm K^{+}}$\, and ${\rm F^{-}}$
ions are approximately equal to a half of the hydrogen bond length
($r_{\rm I} \approx \frac{ 1}{ 2}\, l_{\rm H}$). That is why the
features of their thermal motion considerably depend on the
peculiarities of ion-molecule interaction.

\section{Interpretation of the Obtained Results Based on Microscopic Concepts}

To provide a clearer interpretation of the obtained  results, let us
apply the microscopic approach, in which the interaction of water
molecules and ions is described on the basis of the generalized
Stillinger--David polarization potential {\cite{2,23}} having the
structure
\begin{equation}\label{phiwI}
\Phi_{\mathrm{wI}}=\Phi_{\mathrm{I}}+\Phi_{\mathrm{II}}+\Phi_{\mathrm{III}}+\Phi_{\mathrm{IV}},
\end{equation}
where $\Phi_i$, $i={\rm I, II, III, IV}$ are the components  of the
potential that describe:  $\Phi_{\rm I}$ -- direct Coulomb
interaction with oxygen and hydrogens of a water molecule;
$\Phi_{\rm II}$ -- the charge-dipole interaction between 1) the ion and
oxygen and 2) the ion and hydrogens of a water molecule (it is assumed
that oxygen and the ion acquire a dipole moment due to their
polarization); $\Phi_{\rm III}$ -- interaction of the dipole moments
of oxygen and the ion, and $\Phi_{\rm IV}$ -- repulsion between electron
shells of oxygen and the ion. The explicit form of the contributions
$\Phi_{\rm I}$ --~$\Phi_{\rm IV}$ is given in Appendix.

\begin{figure}
\centering
\includegraphics[natwidth=530, natheight=519, scale=0.33]{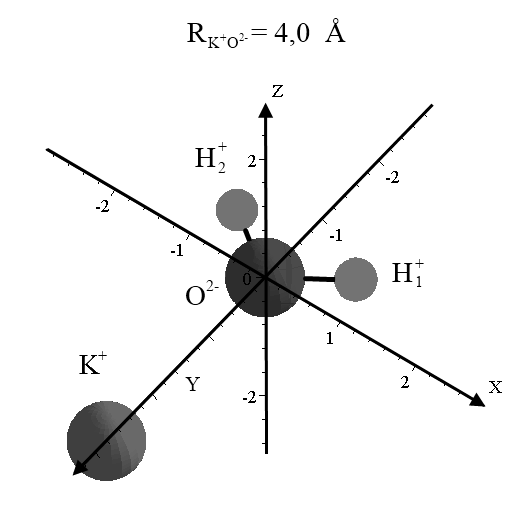}
\caption{Molecular system of coordinates}\label{fi1}
\end{figure}

The positions of ions on equidistant surfaces is determined  by the
polar ($\theta$) and azimuth ($\phi$) angles specified in the
molecular system of coordinates (MSC) (Fig.~1). It is assumed that
oxygen and hydrogens are located in the plane ($x,y$) of the MSC so
that the axis $y$ coincides with the bisector of the angle formed by
$\rm O^{2-}$ and $\rm H_1^{+}$, $\rm H_2^{+}$ and is directed
opposite to hydrogens.

\begin{figure*}
\centering
\includegraphics[natwidth=720, natheight=711, scale=0.25]{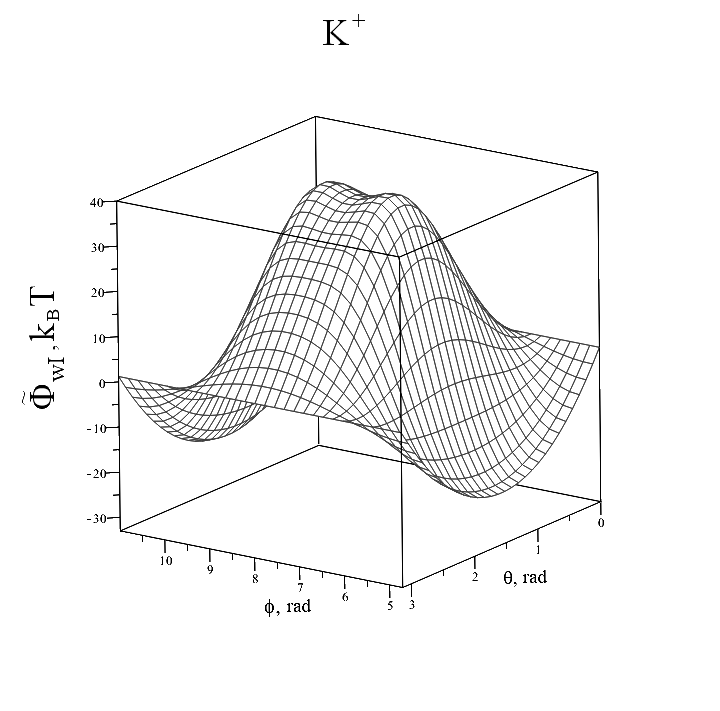}\hspace{1.5cm}\includegraphics[natwidth=720, natheight=712, scale=0.25]{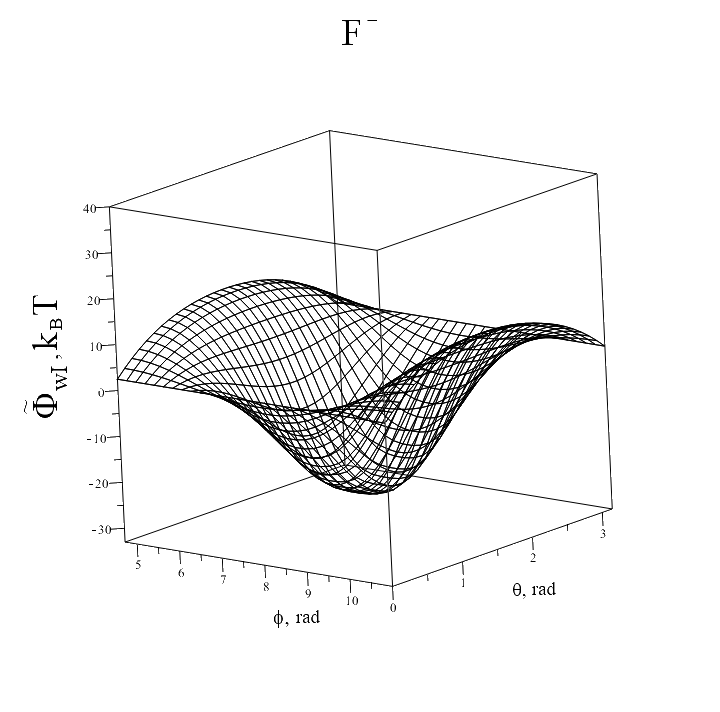}\\
{\large\it ~~~~a\hspace{8.5cm}b} \vskip-3mm \caption{Equidistant
surfaces ($r_{\rm OI} = 4{.}0$ \AA ) of the dimensionless energy of
interaction ${\widetilde {\rm \Phi}} _{\rm wI}$ between a water
molecule and $\rm K^{+}$  ({\it а}), $\rm F^{-}$ ({\it b})
}\label{fi2}
\end{figure*}

The angular dependence of the energy ${\widetilde {\Phi}}
_{\mathrm{wI}}$
 (${\widetilde {\Phi}} _{ \mathrm{wI}} = \left.  {\Phi} _{\mathrm{wI}}
  \right/ k_{\rm B} T_m $, where ${T_m = 273 }$ K is the melting
temperature) of interaction between a water molecule and ${\rm
K^{+}}$, ${\rm F^{-}}$ ions with almost coinciding rigid radii
($r_{\rm c}({\rm K^{+}})=1{.}33 $\, \AA,  $r_{\rm a}({\rm
F^{-}})=1{.}36 $\, \AA \,), is calculated with the help of formula
(\ref{phiwI}) and depicted in Fig. 2. Considerably different
anisotropy characters of the equidistant surfaces of the interaction of
${\rm K^{+}}$ and ${\rm F^{-}}$ ions with a water molecule are
determined by the different structures of their electron shells: for
${\rm K^{+}}$, the shell is argon-like, whereas
it is neon-like for ${\rm F^{-}}$. This is the reason for the noticeably different
polarizations of ${\rm K^{+}}$ and ${\rm F^{-}}$ ions (see Table 3).

At the same time, the situation somewhat simplifies if ions get to
water. This is related to the fact that the interaction of an ion
with surrounding water molecules is not accompanied by significant
changes of the local arrangement of water molecules. Indeed, the
average distance between molecules of water in a vicinity of its
crystallization point is close to 3 \AA, whereas the sum of the
rigid radii of water molecules is approximately equal to the
hydrogen bond length, $l_{\rm H}\approx 2{.}8 $ \AA. So, the volume
of ``empty space'' in water is insignificant. The introduction of
${\rm K^{+}}$ and ${\rm F^{-}}$ ions is actually possible if they
substitute one of the water molecules in local configurations formed, first of
all, by hard cores of the latter. It is worth noting that such
cores are close to spheres to an acceptable accuracy \cite{29,30}.
Hydrogen bonds give rise to considerable dipole correlations, as well
as multipole correlations of higher order. Moreover, structural
violations in water surroundings will grow with increasing
deviations of the cation and anion radii from those of ${\rm K^{+}}$
and ${\rm F^{-}}$. Such structural perturbations of water cannot be
interpreted, if one is based only on concepts of the dominant role of the
hydrogen bond network. However, the analysis based on the averaged
intermolecular interaction between an ion and water molecules is
quite sound. The interaction is naturally averaged due to the
thermal rotation of water molecules. By definition, the averaged
interaction potential $G_{\rm wI}(r)$ is described by the formula
\begin{equation}
G_{\mathrm{wI}}(r)=\frac{\int\int \Phi_{
\mathrm{wI}}(r,\theta,\phi)e^{-\beta\Phi_{\mathrm{wI}}}\sin\theta\,
d\theta \,d\phi}{\int\int e^{-\beta\Phi_{\mathrm{wI}}}\sin\theta\,
d\theta \,d\phi},
\end{equation}
where $\beta=1/k_{\rm B} T$. The averaged interaction energy as  a
function of the distance between ${\rm K^{+}}$ and ${\rm F^{-}}$
ions and water oxygen is presented in Fig.~3, where we also show the
asymptotic behavior of the averaged potential of interaction between
an ion and a water molecule (dotted line) that can be calculated in
the explicit form.

Indeed, at large distances between an ion and oxygen,  the energy of
their interaction is determined by the main terms of the multipole
expansion
\[
\Phi_{\mathrm{wI}}(r)=\frac{q_{\mathrm{I}}
(\textbf{d}_{\mathrm{w}}\cdot \textbf{r})}{r^3}+
\]
\begin{equation}
+\frac{1}{r^3}\left[\textbf{d}_{\mathrm{w}}\cdot\textbf{d}_{
\mathrm{I}}-3\frac{(\textbf{d}_{\mathrm{w}}
\cdot\textbf{r})(\textbf{d}_{
\mathrm{I}}\cdot\textbf{r})}{r^2}\right]+\cdots
\end{equation}
which includes the charge-dipole and dipole-dipole interactions.  Let
the dipole moment of a water molecule be fixed and change only in
direction, while let the ion dipole moment arise due to its polarization
by the electric field of the water dipole:
\begin{equation} \label{dI}
\textbf{d}_{\rm I}=\frac{\alpha_{\rm I}}{r^3}\left[\textbf{d}_{\rm
w} -3\frac{(\textbf{d}_{\rm
w}\cdot{\textbf{r}})\textbf{r}}{r^2}\right].
\end{equation}
Thus, the main contributions to the energy of dipole-ion interaction
take the form:
\begin{equation}\label{PhiwI}
\Phi_{\mathrm{wI}}(r)=\frac{q_{ \mathrm{I}}
(\textbf{d}_{\mathrm{w}}\cdot
\textbf{r})}{r^3}+\frac{\alpha_{\mathrm{I}}}{r^6}\left[\textbf{d}_{
\mathrm{w}}^2 +3\frac{(\textbf{d}_{
\mathrm{w}}\cdot\textbf{r})^2}{r^2}\right]+\cdots .
\end{equation}
At sufficiently large distances between a water molecule and an ion,
\[
\exp{(-\beta\Phi_{\rm wI})}=1-\beta\Phi_{\rm wI}+\cdots.
\]
That is why the asymptotic behavior of $G_{\rm wI}(r)$ will be determined by the contributions
\begin{equation}\label{GwI}
G_{\mathrm{wI}}(r)=\langle \Phi_{
\mathrm{wI}}\rangle_0-\beta\langle\Phi_{
\mathrm{wI}}^2\rangle_0+\cdots ,
\end{equation}
where the index ``0'' denotes the averaging over  the isotropic
orientation distribution of water molecules.

Combining (\ref{dI}), (\ref{PhiwI}), and (\ref{GwI}), it is easy to
make sure that the main contributions to the averaged potential are
determined by the following terms:
\begin{equation}\label{GwI1}
G_{\mathrm{wI}}(r)=-\beta\frac{q_{\mathrm{I}}^2\textbf{d}_{
\mathrm{w}}^2}{r^4}+2\frac{\alpha_{\mathrm{I}} \textbf{d}_{
\mathrm{w}}^2}{r^6}+\cdots .
\end{equation}
One can see that, regardless of the sign of the ion charge, the
averaged interaction between ions and a water molecule at large
distances is of attractive character. For singly charged ions, the
main asymptotic contribution (from the charge-dipole interaction)
does not depend on the type of ion. At the same time, the second
contribution is not universal: it directly depends on the
polarization of ions.

Particularly, Eq. (\ref{GwI1}) together with the analysis of Table 3
make one expecting that, at rather large distances ($r>10 $~\AA)
between an ion and a water molecule, $G_{\rm wF^{-}}(r)>G_{\rm
wK^{+}}(r)$. This conclusion agrees with the calculation results. At
the same time, in the case of relatively small distances ($r<10 $\,
\AA), the averaged potential of interaction between an ${\rm F^{-}}$
anion and a water molecule appears deeper than $G_{\rm wK^{+}}(r)$
(see Fig. 3), owing to which an ${\rm F^{-}}$ anion is less mobile
than a ${\rm K^{+}}$ cation, which also completely agrees with the
data in Table 5.

\begin{figure}
\includegraphics[natwidth=1680, natheight=979, scale=0.3, width=\column]{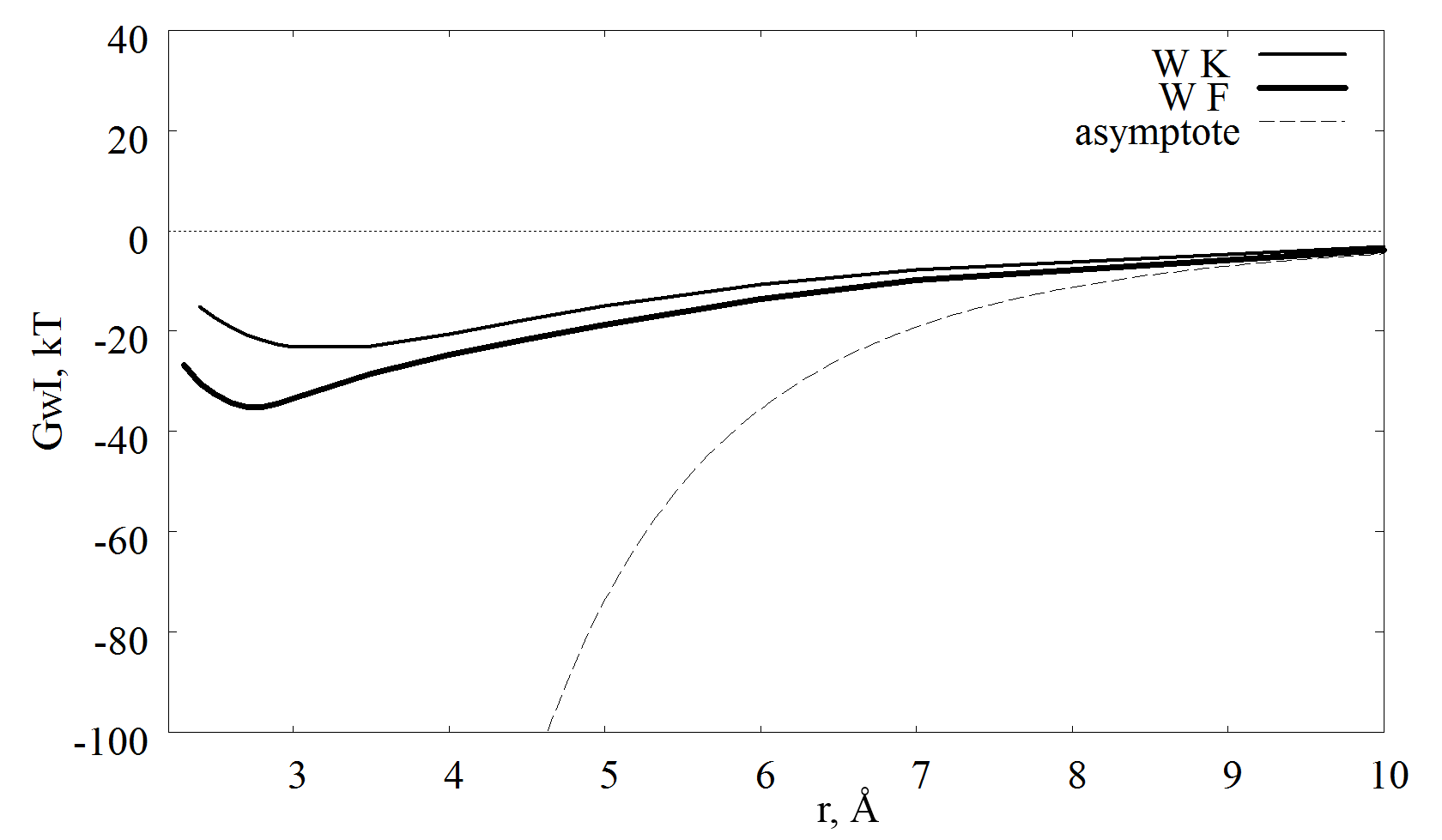}
\vskip-3mm \caption{Distance dependence of the averaged potential
$G_{\rm wI}(r)$
  of interaction of a water molecule with $\rm K^{+}$ ($\alpha = 0{.}83$
  {\AA}$^{3}$) and $\rm F^{-}$ ($\alpha = 1{.}04$ {\AA}$^{3}$) }\label{fi3}
\end{figure}

\section{Discussion of the Results}

The study of the transfer processes in dilute solutions  of
electrolytes performed in this work allows us to make the following
conclusions: 1) the key role in the formation of kinetic properties
of water electrolyte solutions (first of all, in the behavior of the
mobility coefficients of ions and water molecules) is played by
their hard-core radii; 2) the standard understanding of the ion motion
in ``cavities'' formed by the hydrogen bond network is, to our mind,
inconsistent; 3) the hydrogen bond network in water is not
determinative for the character of the ion drift motion, as the
sizes of all cations and anions exceed those of ``cavities'' of the
local water structure. In addition, it should be taken into account
that a certain local configuration of water molecules exists only
during the time coinciding with the duration of the settled life of
molecules ${\tau_0}$ by order of magnitude. Speaking about the role
of hydrogen bonds in the self-diffusion problem, we actually mean
the manifestation of strong orientation correlations between the
dipole moments of water molecules and multipole moments of higher
orders; 4) the behavior of the mobility coefficients of cations and
anions in water agrees with that of the averaged potentials of
interaction between ions and water molecules. Though this fact is
established only for one cation and one anion, it still can be
generalized to all other ions based on the similarity principle.

Analyzing the density of dilute solutions of electrolytes, one can
see that, if one ion is added per approximately 15 water molecules,
the change of the relative volume occupied by one water molecule
does not exceed a half of one per cent. This testifies to the fact that
hydration effects are related, first of all, to the variation of the
local orientation ordering of water molecules rather than to a
local change of the density of the system.

The established connection between the behavior of
cations and anions and the behavior of the averaged potential is a
result of the existence of thermal rotational motion of water
molecules. This fact is most obviously manifested in the temperature
dependences of the dielectric relaxation time and the shear viscosity
of water.

\subsubsection*{APPENDIX}

{\footnotesize Let us consider, in brief, the structure of each
contribution to the energy of interaction between an ion and a water
molecule (Eq. (\ref{phiwI})). The first term $\Phi_{\rm I}$
\cite{1,2,23} determines the direct Coulomb interaction between an
ion (hereinafter denoted by character ${\rm I}$) and oxygen and
hydrogens of a water molecule:
\begin{equation*}
\Phi_{\mathrm{I}}=\sum_{j=0}^{j=2}\frac{q_{\mathrm{I}}\,
q_j}{r_{ij}}, \tag{D1}
\end{equation*}
where $q_{\rm I}$ is the charge of ion ${\rm I}$, $j=0,1,2$, the
index $j$ numbers the charges of oxygen and hydrogen of the water
molecule ($j=0$ corresponds to the charge of oxygen, $j=1,2$ -- to
those of hydrogens). It is assumed that the charges are measured in
units of the electron charge and $q_{\rm I}=\pm n$ ($n$ is the ion
charge), $q_0=-2$, and $q_1=q_2=1$.


\begin{table}[b]

\noindent\caption{Parameters of the potential \cite{2,32} of
interaction  between a water molecule and an ion}
\vskip3mm\tabcolsep12.2pt

\noindent{\footnotesize\begin{tabular}{c c c c c c c c c}
 \hline \multicolumn{1}{c}
{\rule{0pt}{9pt}$ $} & \multicolumn{1}{|c}{$b_3$}&
\multicolumn{1}{|c}{$\rho_3$, {\AA}$^{-1}$}&
\multicolumn{1}{|c}{$b_4$}& \multicolumn{1}{|c}{$\rho_4$,
{\AA}$^{-1}$}\\%
\hline%
$\rm K^{+}$& 64{.}54 & 2{.}569 & 4282{.}4 & 2{.}569 \\%
\hline
{\rule{0pt}{9pt}$ $}& \multicolumn{1}{|c} {$b_3$}& \multicolumn{1}{|c}{$\rho_3$, {\AA}$^{-1}$}& \multicolumn{1}{|c} {$b_4$}& \multicolumn{1}{|c} {$\rho_4$, {\AA}$^{-1}$}\\%
\hline
$\rm F^{-}$& --11{.}45 & 0{.}4304 & 63{.}95 & 1{.}042  \\%
\hline
\end{tabular}}
\end{table}

The second contribution $\Phi_{\rm II}$ determines the potential  of
interaction between the point ion charge and the polarized oxygen of the
water molecule. Oxygen is polarized under the action of the field of
charges of hydrogens in the molecule itself and the charge of the
polarized ion ${\rm I}$. The polarization of oxygen results in the
appearance of the dipole moment $\textbf{d}_{\rm O}$ that
characterizes the degree of deformation of the oxygen electron
shells. The ion is polarized under the action of the hydrogen field
of charges and the charge of oxygen in the water molecule. The
polarization of the ion results in the appearance of the dipole
moment $\textbf{d}_{\rm I}$ characterizing the degree of deformation
of the ion electron shells. With regard for the polarization
contribution, one obtains
\[ \Phi_{\mathrm{II}}   =  \frac{(\textbf{d}_{\mathrm{O}} \cdot
\textbf{r}_{\mathrm{OI}})q_{\mathrm{I}}}{r_{
\mathrm{OI}}^3}[1-L(r_{\mathrm{OI}})]+ \frac{(\textbf{d}_{\rm I}
\cdot \textbf{r}_{\mathrm{IO}})q_{\rm O}}{r_{\mathrm{IO}}^3}[1-L(r_{
\mathrm{IO}})]+
\]
\begin{equation*}
+ \sum_{j=1,2} \frac{(\textbf{d}_{\mathrm{I}} \cdot
\textbf{r}_{{\mathrm{I}}j})q_j}{r_{{\mathrm{I}}j}^3}[1-K(r_{{
\mathrm{I}}j})], \tag{D2}
\end{equation*}
where $\textbf{d}_{\rm O} $ and $\textbf{d}_{\rm I}$ are the dipole
moments of oxygen of the water molecule and the polarized ion ${\rm
I}$, respectively, whereas $1-L(r)$ and $1-K(r)$ are screening
functions (see \cite{1,2,23}). The dipole moment $\textbf{d}_{\rm
O}$ is determined in the molecular system of coordinates, whose
origin coincides with the center of mass of oxygen of the water
molecule. The dipole moment $\textbf{d}_{\rm I}$ is determined in
the molecular system of coordinates, whose origin coincides with the
center of mass of the ion. At large distances, the polarization
contribution $\Phi_{\rm III}$ is reduced to the interaction between
the polarized oxygen of the water molecule and the ion charge, as well as
between the polarized ion and the charges of the water molecule.

The component $\Phi_{\rm III}$ describing the dipole-dipole
interaction  of the ion and oxygen has the form
\begin{equation*}
\Phi_{\mathrm{III}}=\Phi_{\mathrm{d}}(\textbf{d}_{\mathrm{O}}
,\textbf{d}_{ \mathrm{I}})[1 - K({r_{\mathrm{OI}} }/ a)], \tag{D3}
\end{equation*}
where
\[
\Phi_{\mathrm{d}}(\textbf{d}_1,\textbf{d}_2)=\frac{1}{r_{12}^3}\left[\textbf{d}_1\cdot\textbf{d}_2-
\frac{3(\textbf{d}_1\cdot\textbf{r}_{12})(\textbf{d}_2\cdot\textbf{r}_{21})}{r_{12}^2}\right],
\]
and $[1 - K({{r_{\rm OI} } \mathord{\left/
 {\vphantom {{r_{\rm OI} } a}} \right.
 \kern-\nulldelimiterspace} a})]$ is the screening function \cite{2,23}.

The repulsion between the electron shells of the ion and oxygen of
the water molecule is modeled by the exponential functions
\cite{2,23}:
\begin{equation*}
\Phi_{\mathrm{IV}}=\frac{b_3 e^{-\rho_3 r_{\mathrm{IO}}}}{r_{
\mathrm{IO}}}+\sum_{j=1,2}\frac{b_4 e^{-\rho_4 r_{{\rm
I}j}}}{r_{{\mathrm{I}}j}}, \tag{D4}
\end{equation*}
where $b_3$ stands for the amplitude of the ``ion--oxygen''
repulsion energy,  $\rho_3$ is the inverse radius of action of the
repulsion forces between the shells of the ion and oxygen, $b_4$ is
the amplitude of the ``hydrogen--ion'' repulsion energy, and
$\rho_4$ is the inverse radius of action of the repulsion forces
between hydrogens and ion electron shells. Their values are gathered
in Table 9.

At distances significantly exceeding the sizes of a water  molecule
and the ion, the interaction potential takes the asymptote
\begin{equation*}
\Phi  \to \Phi _{\mathrm{d}} ({\textbf{d}_{\mathrm{w}}} ,
{\textbf{d}_{\mathrm{I}}} ), \tag{D5}
\end{equation*}
where ${\textbf{d}_{\rm w}}={\textbf{d}_{\rm H}}+{\textbf{d}_{\rm
O}}$
 is the dipole moment of the water molecule and ${\textbf{d}_{\rm H}}$ is
  the contribution made to this moment by hydrogens.
}

\rezume{%
ОСОБЛИВОСТІ РУХУ КАТІОНІВ І АНІОНІВ\\ В РОЗЧИНАХ ЕЛЕКТРОЛІТІВ}{Л.А.
Булавін, І.В. Жиганюк, М.П. Маломуж,\\ К.М.~Панкратов} {Досліджено
фізичну природу рухливості іонів і молекул води у розбавлених водних
розчинах електролітів, коли на один іон припадає не більше
п'ятнадцяти молекул води. Показано, що поведінка коефіцієнтів
рухливості молекул води і іонів, а також коефіцієнтів самодифузії
молекул води вирішальним чином ви\-значається радіусами їх твердих
оболонок, а не впливом сітки водневих зв'язків у системі.
Встановлено, що вплив гідратацій\-них ефектів на значення густини
системи і коефіцієнтів само\-дифузії молекул води не перевищує
кількох відсотків. На основі мікро\-скопічних уявлень показано, що
відмінна поведінка катіона $\rm K^{+}$ та аніона $\rm F^{-}$, що
мають однакові жорсткі радіуси, добре узгоджується з особливостями
міжмолекулярної взаємодії, яка описується узагальненим потенціалом
Стілінджера--Девіда \cite{1,2}.}

\end{document}